\documentclass[a4paper,conference]{IEEEtran}
\IEEEoverridecommandlockouts
\usepackage{cite}
\usepackage{amsmath,amssymb,amsfonts,amsthm}
\usepackage{newtxmath}
\usepackage{algorithmic}
\usepackage{graphicx}
\usepackage{booktabs}
\usepackage{textcomp}
\usepackage{xcolor}
\usepackage{soul}
\def\BibTeX{{\rm B\kern-.05em{\sc i\kern-.025em b}\kern-.08em
    T\kern-.1667em\lower.7ex\hbox{E}\kern-.125emX}}

\theoremstyle{definition}

\usepackage{graphicx}

\usepackage[hyphens]{url}
\usepackage{breqn}

\usepackage{caption}
\usepackage{subcaption}

\begin{document}

\title{Vision-Based Reconfigurable Intelligent Surface Beam Tracking for mmWave Communications
\thanks{Special thanks to the Sony Research Center in Lund for providing their reconfigurable intelligent surface for testing and research.

This work has been funded by the Horizon Europe EU Framework Programme under the Marie Skłodowska-Curie grant agreement No.\,101059091, the Horizon 2020 EU Framework Programme under Grant Agreement No.\,861222, the Swedish Research Council (Grant No. 2022-04691), the strategic research area ELLIIT, Excellence Center at Linköping – Lund in Information Technology, and Ericsson.
}
}

\author{
\IEEEauthorblockN{Juan Sanchez, Xuesong Cai, and Fredrik Tufvesson}
\IEEEauthorblockA{Department of Electrical and Information Technology, \textit{Lund University}, Lund, Sweden \\
\{juan.sanchez, xuesong.cai, fredrik.tufvesson\}@eit.lth.se }

}

\maketitle

\begin{abstract}

Reconfigurable intelligent surfaces have emerged as a technology with the potential to enhance wireless communication performance for 5G and beyond. However, the technology comes with challenges in areas such as complexity, power consumption, and cost. This paper demonstrates a computer vision-based reconfigurable intelligent surface beamforming algorithm that addresses complexity and cost issues and analyzes the multipath components that arise from the insertion of such a device into the wireless channel. The results show that a reconfigurable intelligent surface can provide an additional multipath component. The power of this additional path can be critical in blockage scenarios, and a capacity increase can be perceived in both line-of-sight and non line-of-sight scenarios.


\end{abstract}

\section{Introduction}
\label{sec:introduction}

The ever-increasing need for higher data rates and the scarcity of the radio spectrum have been influential factors for the exploration of communication at higher frequencies with each new technology release. Propagation at radio bands such as millimeter wave (mmWave) or subterahertz starts to show characteristics typical of optical propagation, where line-of-sight (LOS) and specular reflections are the dominant propagation mechanisms \cite{haneda2015statistical,fan2016measured}. Propagation at these frequencies also suffers from higher propagation loss and higher absorption effects. Massive multiple-input multiple-output (MIMO) systems have exploited a large number of antenna elements to enable coherent processing of electromagnetic waves (EM) and withstand the harsh wireless channel at higher frequencies at the transmitting and receiving ends.

Recently, the paradigm has further evolved into a joint manipulation of the EM waves from the transmitting and receiving ends and from the environment itself, by using so-called reconfigurable intelligent surfaces (RISs). These RISs are devices that are thought to scatter incident EM waves at will depending on the desired configuration. The technology comes with practical challenges such as manufacturing cost, power consumption, processing complexity -- which in turn affects delay and power consumption, among all. Multiple studies have shown that RISs can leverage the geometry of nodes and the environment to form a beam at low complexity. For this purpose, low-cost optical devices such as cameras can be used as sensors whose output can be processed using computer vision algorithms, revealing the geometry of the environment. The authors in \cite{jiang2023sensing,ouyang2023computer} demonstrated that the geometric information extracted from cameras can improve the SNR and consequently the achievable bitrate by using different processing methods.

The present paper demonstrates an arrangement of a computer vision-aided RIS that tracks the RX and TX location and updates its scattering pattern according to the beamforming algorithm presented in \cite{sanchez2021optimal}. Furthermore, the effect of introducing such a structure into the wireless channel is analyzed. The rest of the paper is organized as follows: Section \ref{section:measurementCampaign} presents the RIS arrangement and measurement setup. Section \ref{section:mpcAndCapacity} establishes the post-processing and measurement analysis framework. Section \ref{section:resultsDiscussion} presents the measurement results for the double-directional channel impulse response with and without RIS activation and parameterizes the wireless channels experienced in the measurements for each scenario. Section \ref{section:conclusions} gives conclusive remarks and outlook for future studies.

\section{Measurement Campaign}
\label{section:measurementCampaign}

\begin{figure}
\begin{center}
	\includegraphics[width=0.8\columnwidth]{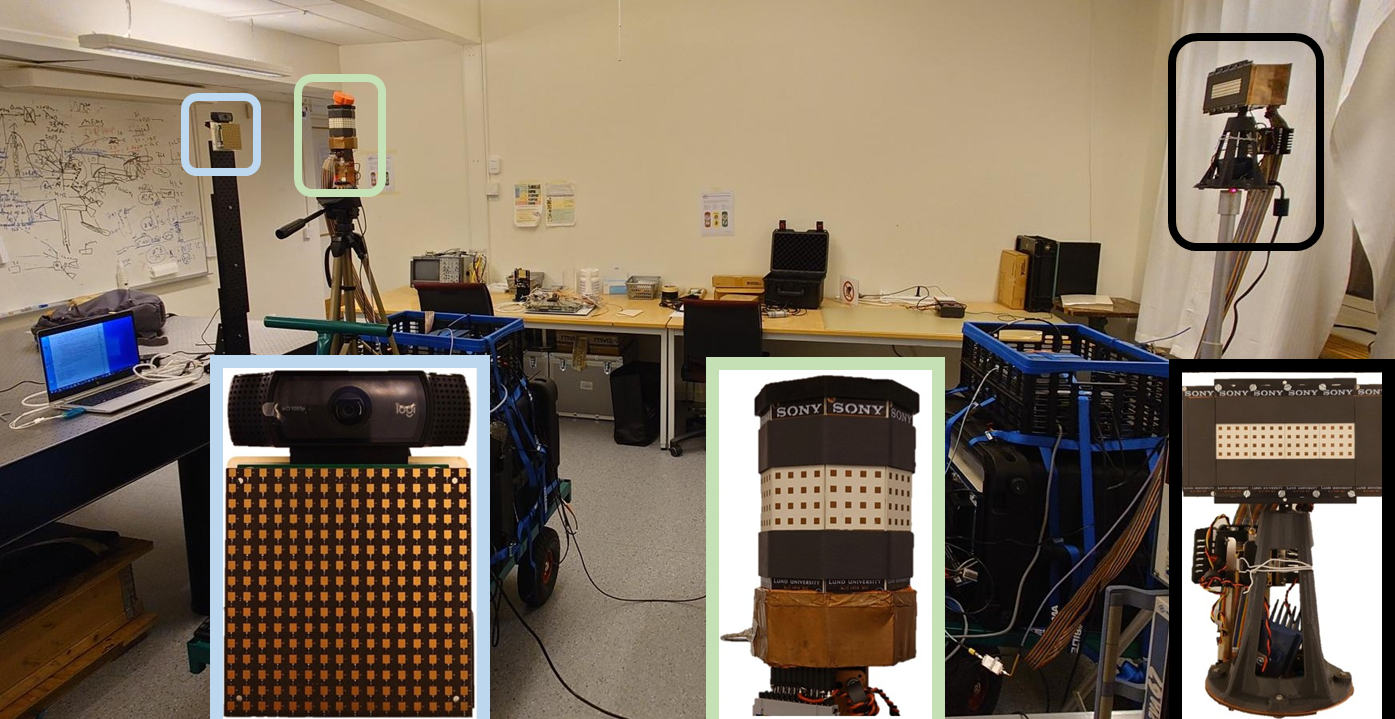}
	\caption{mmWave setup and components. TX array (left), RX Array (center) and RIS with camera on top (right).}
        \vspace{-5mm}
	\label{fig:measurementSetup}
\end{center}
\end{figure}

For the measurement campaign, the mmWave channel sounder available at Lund University \cite{cai2023switched} was used together with a commercial RIS \cite{yezhen2020novel}, in an indoor scenario. The channel sounder measures the channel transfer functions (CTFs) of all antenna combinations available in a so-called snapshot. The channel sounder RF ends consisted of a rectangular TX array and an octagonal RX array, both dual polarized. There were 128 TX and 256 RX antenna elements, for a total of 32768 channels measured per snapshot. The RIS consisted of a 1-bit surface arranged in 16x16 antenna patches, which can be reconfigured via serial communication with another device. A form of geometry-based maximum ratio transmission (MRT) beamforming \cite{sanchez2021optimal} was used to control the device's phase shifts upon visual input from a commercial, low-cost optical camera. Fig.\,\ref{fig:measurementSetup} shows the measurement setup and close-ups of the mmWave components used for this measurement campaign, while Table \ref{tab:systemPar} summarizes all relevant system parameters.

\begin{table}
\begin{center}
	\caption{Parameter setup.}
	\label{tab:systemPar}
	\begin{tabular}{|c|c|}
		
		\hline		
		\textbf{Carrier frequency}	& $28$\,GHz\\\hline
		\textbf{Signal bandwidth}	& $768$\,MHz\\\hline
		\textbf{Delay resolution} & $1.3021$\,ns ($39.04$\,cm)\\\hline
		\textbf{Switching time} & $18.8$\,$\mu$s\\\hline
		\textbf{Snapshot time} & $~600$\,ms\\\hline
		\textbf{Trajectory length} & $1$\,m\\\hline
		\textbf{Total number of snapshots} & $128$\\\hline
	
	\end{tabular}
	\vspace{-4mm}
\end{center}
\end{table}

The RX was moved along a 1\,m straight trajectory normal to the RIS boresight at a constant speed of 1\,cm/s, in a line-of-sight (LOS) scenario, as seen in Fig.\,\ref{fig:measurementLayout}. The trajectory included a number of static measurements at the beginning and end of the trajectory, meaning that the experienced propagation channel should not vary significantly for these positions. This observation implies that the number of available positions is larger than the expected 100 positions for such a trajectory and speed, and that these static measurements are used to validate the consistency in the measurement campaign. The same trajectory was measured for subscenarios where the RIS was active and inactive. The shadowed region in the top left corner of Fig.\,\ref{fig:measurementLayout} corresponds to a virtual blocker added after in the middle of the processing to generate a synthetic non line-of-sight (NLOS) scenario, as discussed in Section \ref{section:mpcAndCapacity}. Calibration was performed in the position axis to align the trajectories for an active and an inactive RIS after measurements were taken.

\section{Multipath Component Extraction and Capacity Analysis}
\label{section:mpcAndCapacity}

The propagation channel can be characterized by a transfer function. The channel transfer function, in turn, is considered as a superposition of multipath components (MPCs) modeled as follows,
\begin{equation}
\label{eq:ctfMPC}
\begin{split}
& \mathbf{H}(f,s,m_{\text{T}},m_{\text{R}};\mathbf{\Theta})\\ & = \sum_{l=1}^L \mathbf{b}_{m_{\text{R}}}^T (\phi_{\text{R},l},\varphi_{\text{R},l},f)
\begin{bmatrix}
    \gamma_{\text{HH},l} & \gamma_{\text{HV},l} \\
    \gamma_{\text{VH},l} & \gamma_{\text{VV},l} 
\end{bmatrix}
\mathbf{b}_{m_{\text{T}}}^T (\phi_{\text{T},l},\varphi_{\text{T},l},f) \\
& \cdot \mathbf{b}(f) e^{-j2 \pi f \tau_l} e^{ -j2 \pi f \nu_l t_{s,m_{\text{T}},m_{\text{R}}} } + \mathbf{N}(f,s,m_{\text{T}},m_{\text{R}}),
\end{split}
\end{equation}
where $\mathbf{H}$ is the CTF dependent on frequency $f$, snapshot index $s$, transmit/receive antenna indices $m_{\text{T}}$, $m_{\text{R}}$, and structural parameters of the propagation channel $\mathbf{\Theta}$ such as directions $\phi$ and $\varphi$, delay $\tau$, Doppler $\nu$ and polarimetric coefficients $\pmb{\gamma}$. Furthermore, $\mathbf{b}_{m_{\text{T}}},\mathbf{b}_{m_{\text{R}}}$ are mappings from departure / arrival directions to polarimetric antenna responses, for transmit and receive antenna elements $m_{\text{T}},m_{\text{R}}$, respectively. In turn, $\mathbf{b}(f)$ is the mapping from frequency to polarimetric responses of the antenna elements. $\mathbf{N}$ denotes white Gaussian noise. The CTF can also be seen as the Fourier transform of the channel impulse response.

An implementation of the space-alternating generalized expectation maximization algorithm (SAGE) \cite{HRPEcai, fleury1999channel} was used to detect the MPCs present at each measured position and estimate their parameters. Using the double-directional polarimetric capabilities of the channel sounder \cite{cai2023switched} and the knowledge of the effective aperture distribution function (EADF) of the sounder antenna arrays \cite{cai2023enhanced}, it is possible to extract a very accurate representation of the characteristics of the propagation channel that is independent of the antenna architectures.

\begin{figure}
\begin{center}
	\includegraphics[width=0.7\columnwidth]{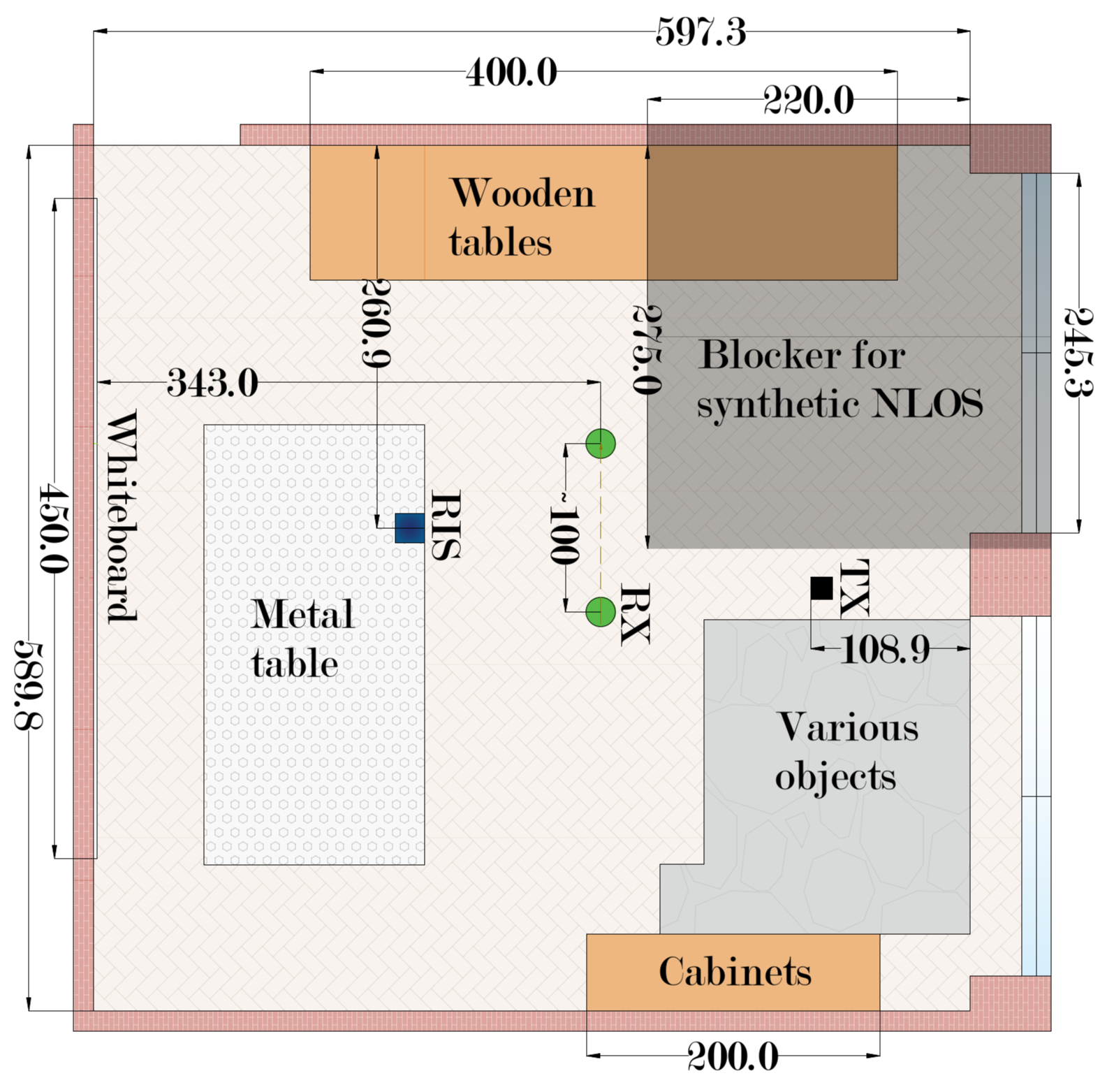}
	\caption{Measurement layout (measures in cm).}
        \vspace{-4mm}
	\label{fig:measurementLayout}
\end{center}
\end{figure}

\begin{figure}
\begin{center}
	\includegraphics[width=0.92\columnwidth]{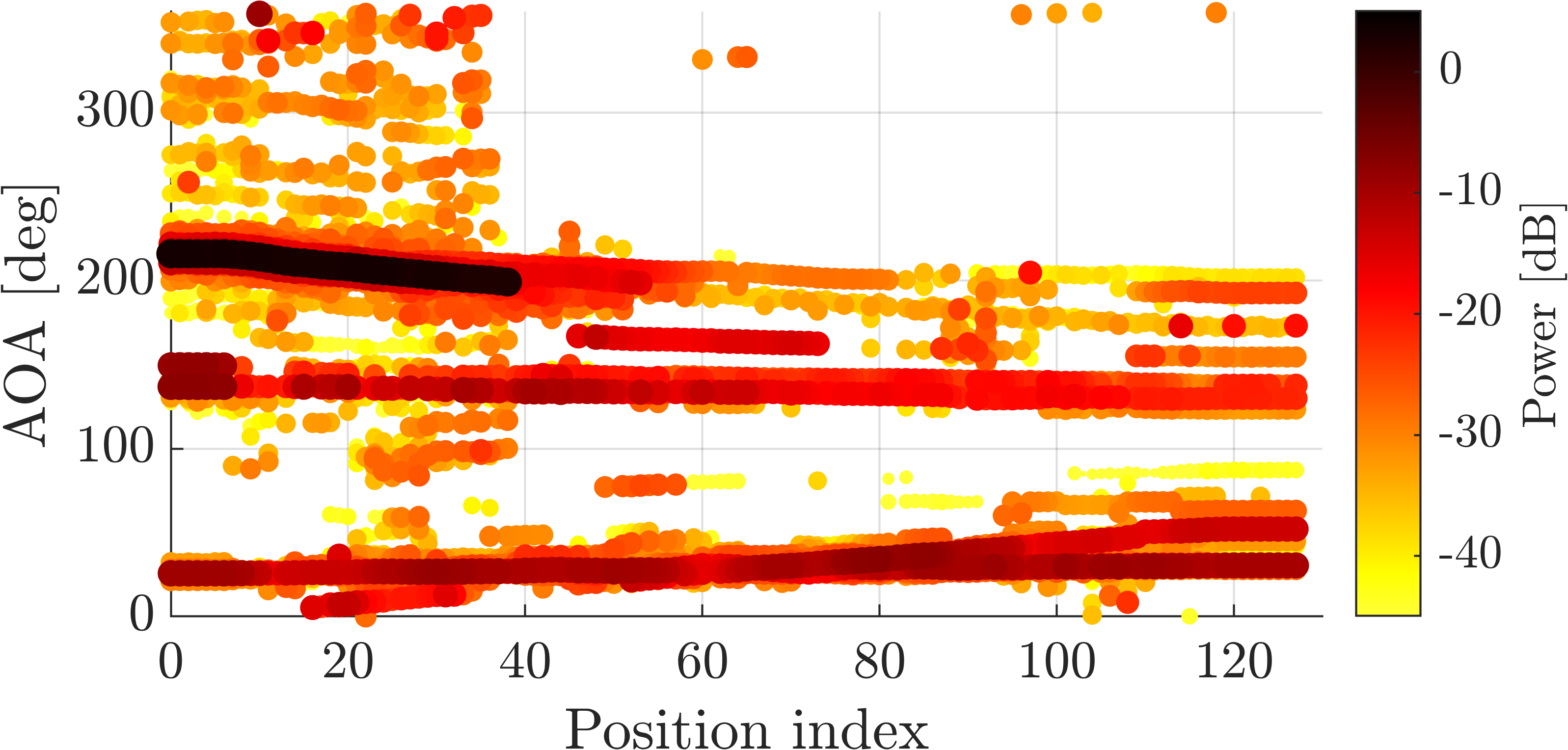}
	\caption{Estimated AOA for the MPCs with active RIS, LOS region (before position 40) and NLOS region (after position 40).}
        \vspace{-8mm}
	\label{fig:losAndNLOSRegions}
\end{center}
\end{figure}

An additional synthetic NLOS region was generated after the estimation of the MPCs, for the assessment of the potential of the RIS technology under different conditions. The synthetic scenario considered an additional blocker shadowing the regions of the measurement layout shown in Fig.\,\ref{fig:measurementLayout}. The insertion of blocking was virtually performed by removing the MPCs with associated angular departure or arrival values that would impinge on the hypothetical surface. This means that the virtual blocker was assumed to be a perfect absorber. The synthetic scenario relied on equation (\ref{eq:ctfMPC}) -- with a reduced number of MPCs -- for the reconstruction of the CTF. Fig.\,\ref{fig:losAndNLOSRegions} shows the stark contrast in the transition from a LOS to a NLOS scenario, which is an approximation of the smoother transition expected in a real scenario. The transition from LOS to NLOS is completed at position 40, implying that the NLOS scenario was bounded from position 40 and onward. Qualitative analysis of the estimated MPCs, as well as quantitative analysis of power and capacity values, were performed to assess the RIS capabilities in the LOS and NLOS scenarios. Since the RIS is, among other things, composed of a metal ground plane, a comparison between an active and an inactive RIS is equivalent to a comparison between a RIS and a metal plate located in the same position.

The effect of introducing a RIS in the environment can be characterized in power by the ratio
\begin{equation}
\label{eq:powerRatio}
    \Delta_P = \frac{||\mathbf{H}_{\text{on}}||^2}{||\mathbf{H}_{\text{off}}||^2} = \sum_{l=0}^L \frac{||\pmb{\gamma}_{l,\text{on}}||_F^2}{||\pmb{\gamma}_{l,\text{off}}||_F^2},
\end{equation}
where $\Delta_P$ is the ratio between the experienced channel gain for a RIS-enabled setup with respect to a scenario with a metal plate in place, $\mathbf{H}$ is the CTF from (\ref{eq:ctfMPC}), and $||\pmb{\gamma}_l||_F$ is the Frobenius norm of the polarimetric matrix for the \textit{l}th channel path. Since the RIS used in this study was found to have polarization-dependent performance variations, we only considered the vertical to vertical polarization pair as a propagation mechanism and left the rest of the polarization pairs for further study. Therefore, (\ref{eq:powerRatio}) can be simplified to
\begin{equation}
\label{eq:powerRatioSimplified}
    \Delta_P = \sum_{l=0}^L \frac{|\gamma_{\text{VV},l,\text{on}}|^2}{|\gamma_{\text{VV},l,\text{off}}|^2},
\end{equation}
where $|\gamma_{\text{VV},l,\text{on}}|^2$ is the squared magnitude of the vertical-vertical polarization coefficient of the channel for the \textit{l}th path. $\Delta_P$ was calculated for each position across the RX trajectory, and a moving average filter of 5 position samples was implemented on $\Delta_P$ to filter out noise-like fast variations and better visualize the tendency of the channel gain ratio.

For the calculation of capacities, it was assumed that channel state information was known at the transmitter (CSIT). Studies in time \cite{caire1999capacity,vu2007mimo,you2020spectral} and literature \cite{molisch2012wireless} have shown that optimal power allocation and consequent capacity maximization can be achieved by using the water filling algorithm. The achieved capacity per bandwidth unit can be computed as
\begin{equation}
\label{eq:csitCapacity}
    C = \sum_{k=1}^{R_{\text{H}}} \log_2 \left( 1 + \frac{P_k}{\sigma_n^2} \sigma_k^2 \right),
\end{equation}
where $R_{\text{H}}$ is the rank of the CTF $\mathbf{H}$, $\sigma_n^2$ is the noise power density, $\sigma_k^2$ is the \textit{k}th singular value of $\mathbf{H}$, and $P_k$ is the power allocated to the \textit{k}th eigenmode of the channel satisfying the constraint $\sum P_k = P$, with $P$ being the total power allocated to a single subcarrier across all channel eigenmodes. The noise spectral density $\sigma_n^2$ was normalized to 0\. dB/Hz. The total power $P$ was defined to be the same across all frequencies, and dictated by signal-to-noise ratio (SNR) values ranging from -10\,dB to 30\,dB. Capacities were calculated for every frequency bin within the sounder's bandwidth, every SNR value, and every trajectory position, and mean and standard deviation statistics were extracted across frequency. In addition, the ratio between the mean capacities for active and inactive RIS was calculated for each position and the SNR value. Finally, for both active and inactive RIS, the mean capacities are further averaged across the positions belonging to the LOS or NLOS scenarios and plotted against the SNR evaluation range.

\section{Results and Discussion}
\label{section:resultsDiscussion}

\subsection{Multipath Component Analysis}

\begin{figure}
    \centering
    \begin{subfigure}[b]{\columnwidth}
        \centering
        \includegraphics[width=\columnwidth]{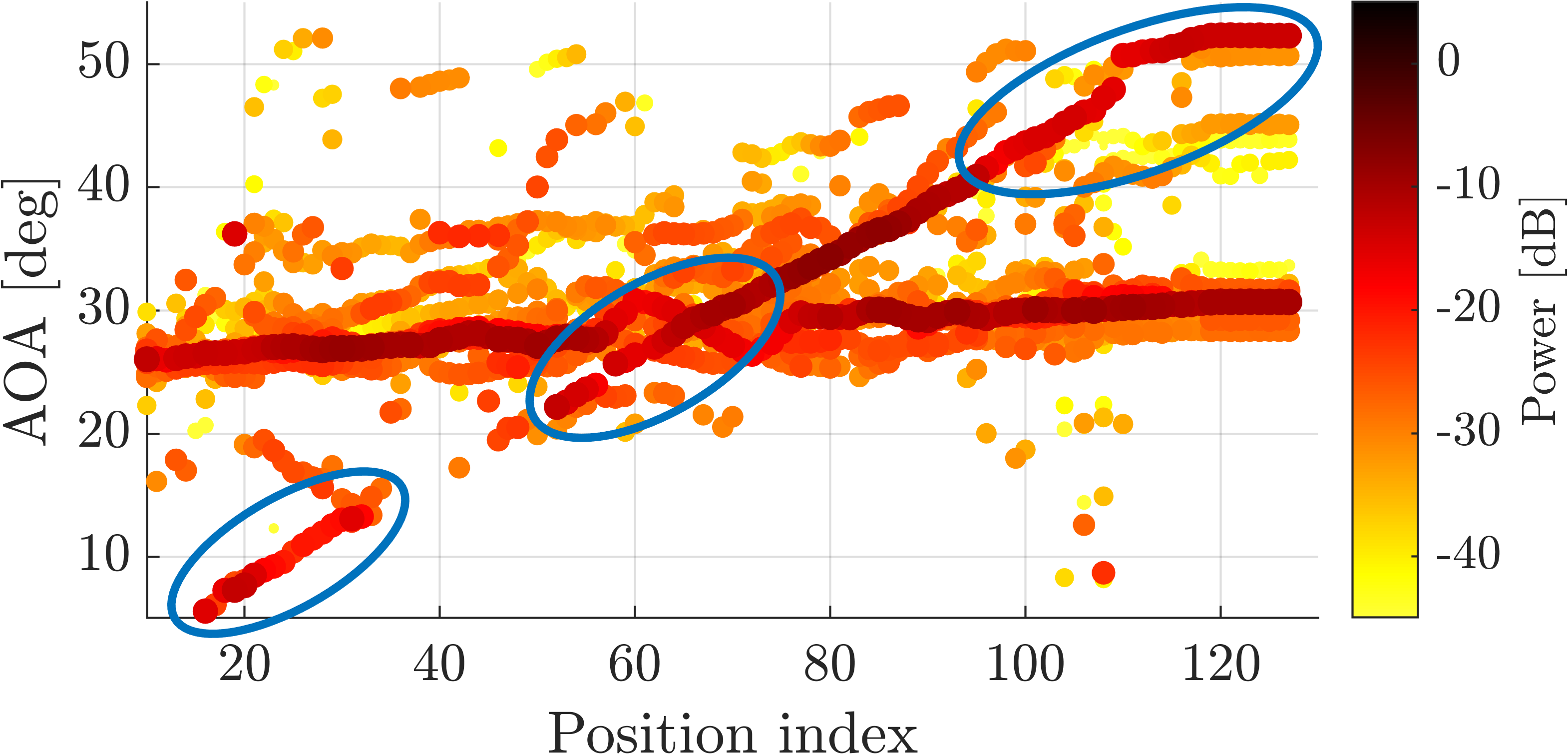}
        \caption{Active RIS.}
        \label{fig:aoaRISZoomOn}
    \end{subfigure}
    \hfill
    \begin{subfigure}[b]{\columnwidth}
        \centering
        \includegraphics[width=\columnwidth]{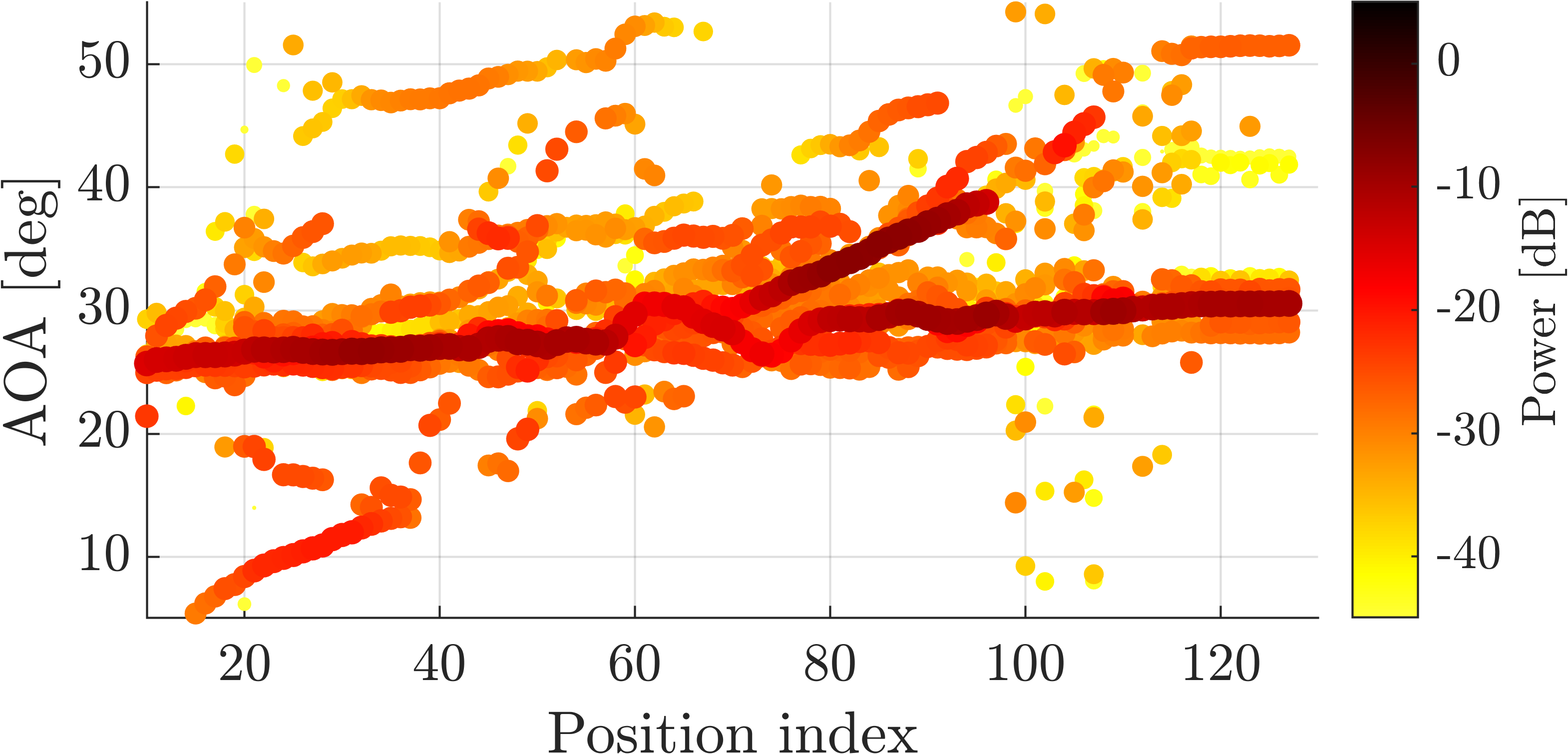}
        \caption{Inactive RIS.}
        \label{fig:aoaRISZoomOff}
    \end{subfigure}
    \caption{Close-up of the RIS effect in the AOA range, with enhancements of MPCs being marked in (a).}
    \vspace{-4mm}
    \label{fig:aoaRISZoom}
\end{figure}

Fig.\,\ref{fig:losAndNLOSRegions} shows the estimated received power and azimuth of arrival (AOA) values of each detected MPC at all positions. The behavior of the entire surface plot remains fairly constant for some few first and last measurement instances, validating the consistency of the measurement results for stationary positions at the beginning and end of the trajectory. It is clear that the strongest MPC that starts around 220$^{\circ}$ corresponds to the LOS path between the TX and the RX. This path decreases in angle along with the movement of the RX across the measurement trajectory, and disappears when transitioning from the LOS to the NLOS region. The MPC cluster correlated in azimuth with the LOS path also gradually disappears. As expected for the LOS region, rich scattering is observed across the whole angular domain, with numerous MPCs coming from various elevation angles when further evaluated in the rest of the directional domains. The MPC cluster with AOA values just below 150$^{\circ}$ corresponds to reflections from the cabinets in Fig.\,\ref{fig:measurementLayout}. The evolution of these MPCs shows power-level variations correlated with the scattering characteristics of the different components in the cabinets, mainly wood and glass. Since these reflections are coming from a direction close in angle with the direction of movement of the RX, their AOA remains relatively constant with respect to other MPCs observed in the measurement. Finally, there is a second most dominant MPC cluster that approximately varies from 0$^{\circ}$ to 60$^{\circ}$. This cluster corresponds to reflections coming from both the laboratory whiteboard and the RIS. As this cluster is of special importance for the purpose of this paper, let us take a closer look at the evolution of these MPCs, and foremost, at the comparison between the active and inactive RIS scenarios.

Fig.\,\ref{fig:aoaRISZoom} shows a close-up of the MPC estimates for the AOA range of interest, for both active and inactive RIS scenarios. In both cases, we observe an MPC cluster with AOA values approximately ranging from 25$^{\circ}$ to 30$^{\circ}$, corresponding to reflections from the whiteboard. As the whiteboard is made up of metal, the reflections coming from this cluster have a significant contribution at the power level. Moreover, the reflections from the whiteboard are not affected by the blocker introduced into the layout and remain independent of the visibility region in which the RX is located. The same can be said of the contribution from the RIS MPC, seen in Fig.\,\ref{fig:aoaRISZoomOn} as the relatively straight line ranging from 5$^{\circ}$ to 52$^{\circ}$ in AOA and "crossing" the MPC cluster coming from the whiteboard, and as the rather shorter/interrupted line in Fig.\,\ref{fig:aoaRISZoomOff} mainly with power contributions from 32$^{\circ}$ to 40$^{\circ}$ in AOA. As the RIS in its inactive state behaves as a metal surface, reflection occurs as a natural process for the latter AOA interval. Even though the area of the RIS is much smaller than the area where reflections from the whiteboard occur, the RIS is closer to the RX and the wavefront travels a shorter distance until it reaches the RX. This results in less spreading and lower pathloss, which in turn implies that the received power is higher than for longer path distances. The AOA values for the whiteboard's MPC cluster increase along with the trajectory except for some noticeable variations seen approximately between positions 60 and 80. These variations can be explained by the geometry of the measurement trajectory: At a certain position in the trajectory, the MPCs coming from the whiteboard and the RIS align, but the RIS is closer to the RX, hence it blocks the radiation coming from the whiteboard. However, diffraction happens around the edges of the RIS, which changes the AOA with which the RX receives some of the whiteboard's MPCs with respect to the expected AOA for no obstruction from the RIS. Hence, the oscillation around the straight linear evolution of the AOA and the reduction in received power for these MPCs is, at a level where the RIS MPC becomes the most dominant among all MPCs in the environment.

Looking further at the RIS MPC, it is very clear that the MPC's angular range for the active RIS is greatly extended, to around 400\% of the angular range of a  metal surface of the same dimensions. The straight evolution line of the MPC is interrupted between 15$^{\circ}$ and 25$^{\circ}$ in AOA. This artifact is a manifestation of the blockage that the RX performs when crossing the path between the TX and the RIS on its trajectory.

\begin{figure}
\begin{center}
	\includegraphics[width=0.95\columnwidth]{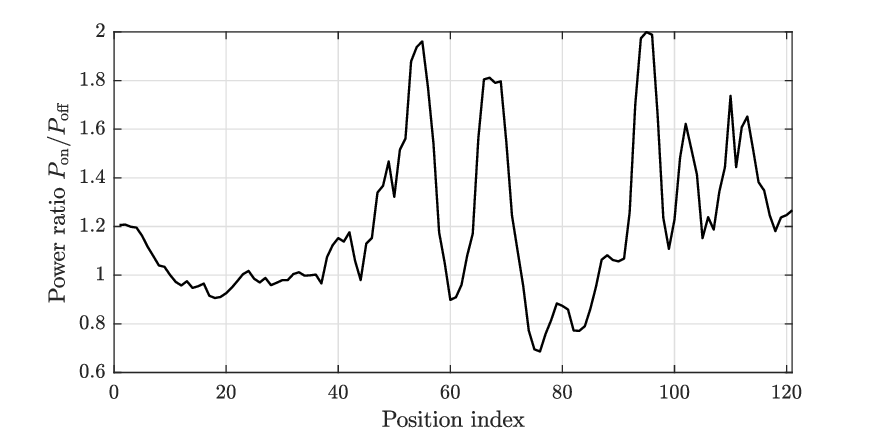}
	\caption{Received power ratio between active and inactive RIS scenarios.}
        \vspace{-6mm}
	\label{fig:powerRatio}
\end{center}
\end{figure}

Let us now look at how the power relations between MPCs in the different scenarios across the visibility regions. Fig.\,\ref{fig:powerRatio} shows the received power ratio between all MPCs for the active RIS with respect to the inactive RIS. There is a slight improvement of the power ratio at the beginning of the trajectory followed by a rather unitary relation for the remaining positions belonging to the LOS region. This relation shows that the power enhancement that the RIS brings in a small indoor LOS environment is small, but noticeable in this setup. In contrast, the improvement induced in the NLOS region of the measurements is much higher and stays for the major part of the trajectory over 1, with most of the peaks above 1.4 and going up to 2 in the best case. The overall improvement corresponds to logarithmic gains of up to 0.8\,dB for the LOS region and up to 3\,dB for the NLOS region in our setup. Notice that for the trajectory around position 80 we see a drop in the ratio, going as low as 0.7. Two factors are involved in this result. The first factor is the natural reflection of the RIS that occurs in this position range for the scenario with inactive RIS. The only gain from an active RIS would be a focusing effect, which is not as pronounced because the beamforming algorithm only took an approximate value for the distance between the RX and the RIS for the whole trajectory. Clearly, for such a trajectory, the distance between the RX and the RIS changes, and beamforming loses focus. The use of multiple optical camera lenses is expected to increase the beamforming gain by accurately estimating the distance to the RX. The second factor is the absence of the strongest MPC that comes from the reflection of the whiteboard at position 82 in Fig.\,\ref{fig:aoaRISZoomOn}, which reduces the entire power contribution for the active RIS scenario. After a closer inspection, some disturbances to the smooth evolution of the mentioned MPC can be seen around the area, which happens to be very close to the area where the whiteboard reflection impinges on the RIS. Further studies would be needed to assess whether the activation of the RIS is playing a role in the disturbances observed for the whiteboard's MPC cluster.

\subsection{Capacity Analysis}

\begin{figure}
\begin{center}
	\includegraphics[width=0.8\columnwidth]{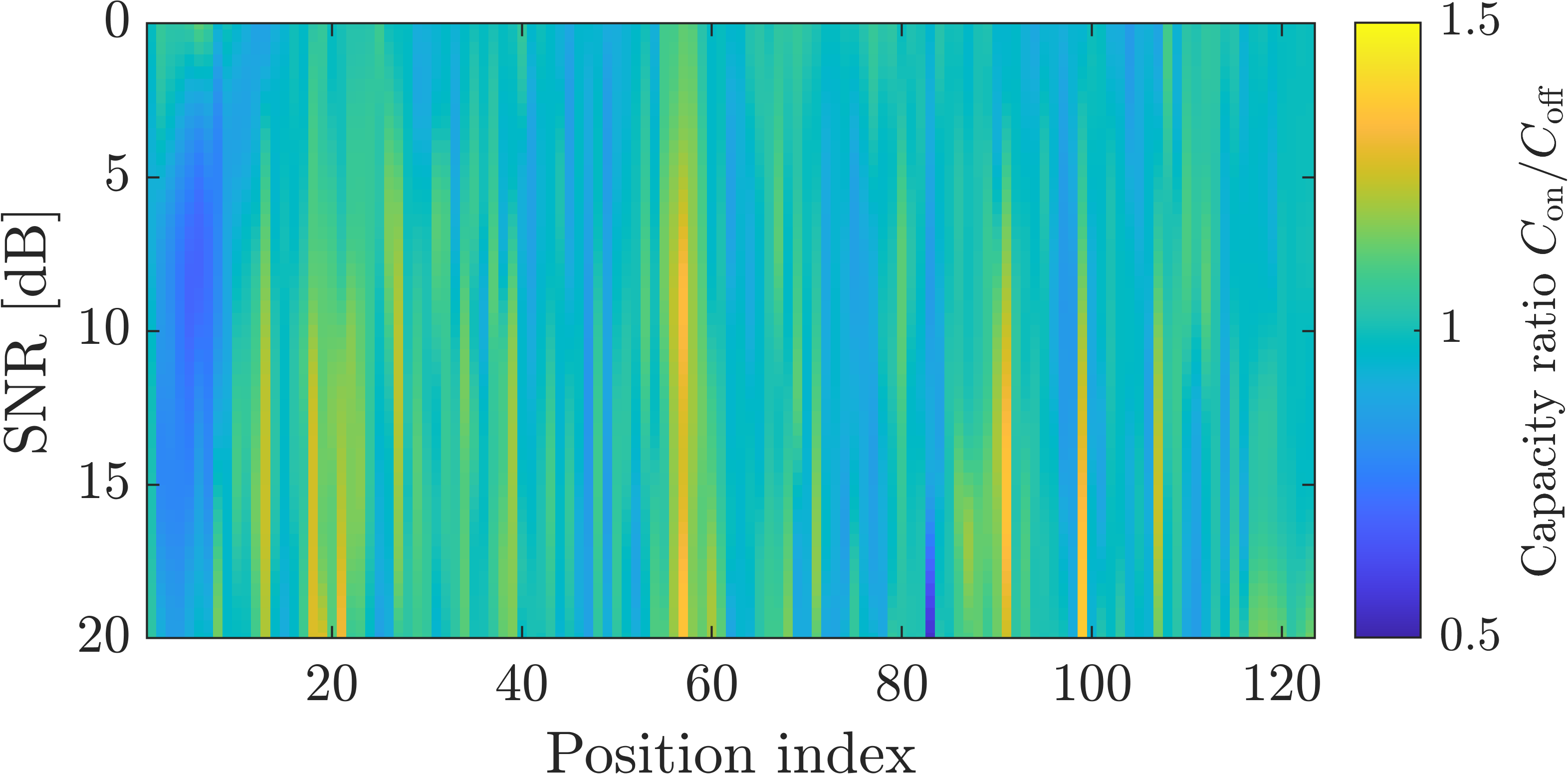}
	\caption{Ratio between capacities with active and inactive RIS.}
        \vspace{-4.5mm}
	\label{fig:capacityRatio}
\end{center}
\end{figure}

For the capacity analysis, a normalization of the CTFs across frequency bins and antenna indices was carried out, thus removing the additional gain from the active RIS seen in Fig. \ref{fig:powerRatio} and allowing an analysis of the small-scale effects of the RIS. Fig.\,\ref{fig:capacityRatio} shows the mean capacity across frequency points, for different positions and SNR values. For low SNR values the ratio remains relatively unitary, but as the SNR increases, there are more pronounced variations in the regions where the ratio is greater or less than 1. There is correspondence between the increases in mean capacity around positions 20, 60, 90 and 120, and the extension of the RIS MPC when going from an inactive to an active state in Fig.\,\ref{fig:aoaRISZoom}.

\begin{figure}
\begin{center}
	\includegraphics[width=0.8\columnwidth]{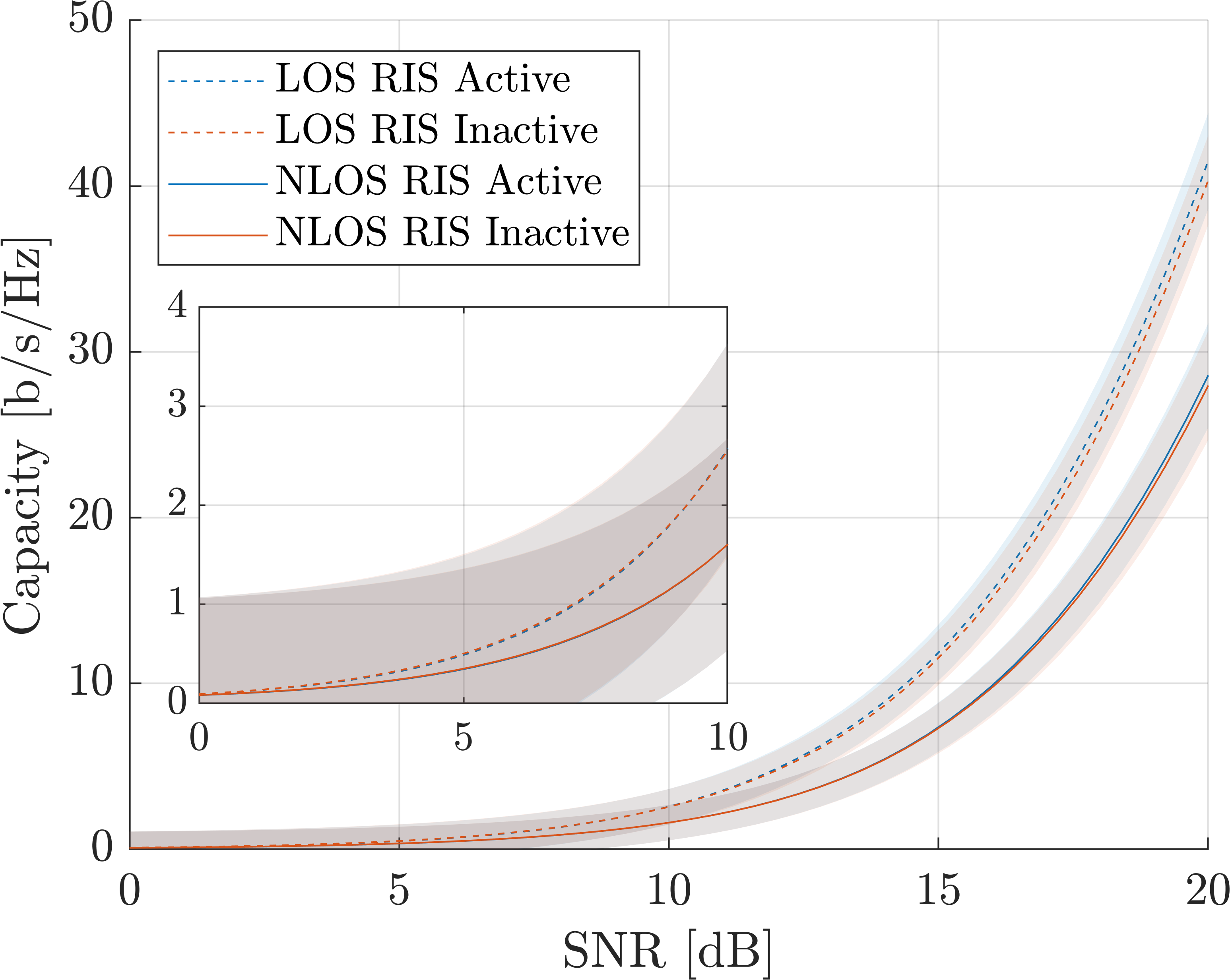}
	\caption{Mean capacity across all positions for all scenarios. The shadowed regions represent the mean standard deviation of the capacity curves.}
        \vspace{-7mm}
	\label{fig:meanCapacities}
\end{center}
\end{figure}

Furthermore, averaging the capacities on all the positions belonging to a certain visibility region, Fig.\,\ref{fig:meanCapacities} shows the mean capacity between frequencies and positions for different SNR values, visibility regions, and active/inactive RIS scenarios. The overall gap between mean capacities for the LOS and NLOS regions is very clear, whereas the gap between mean capacities of active and inactive RIS scenarios is less evident, but still noticeable. An evaluation of the relative ratio between capacities shows that there are mean relative capacity increases of 0.29\% and 0.00\% and maximum relative capacity increases of 3.21\% and 2.20\% for the active RIS with respect to the inactive RIS across the entire SNR range, for the LOS and NLOS regions, respectively. This indicates that although the RIS influences a greater overall power increase in the NLOS scenario, its use to slightly enhance capacity in mmWave communications has a similar effect for both LOS and NLOS scenarios.

\section{Conclusions}
\label{section:conclusions}

This paper showed a detailed analysis of the detected multipath components (MPCs) of a measured indoor scenario with a virtual blocker for the line-of-sight (LOS) and non-LOS (NLOS) regions, where a reconfigurable intelligent surface (RIS) performing beam tracking sought to enhance communication at mmWave frequencies. Beam tracking was performed under a low-complexity geometry-based beamforming algorithm that exploits the angular and range information obtained by visual sources -- a low-cost optical camera in this case. The measurement campaign considered positions across a straight trajectory where two scenarios were depicted, namely, one with an active RIS and the other with an inactive RIS, equivalent to placing a metal surface of the same dimensions. The results show that the activation of the RIS induces a significant extension of the visibility of the associated MPC over the AOA range, to the point where the RIS component becomes the dominant MPC for propagation for some positions in the trajectory. Furthermore, measurements show that the inclusion of a RIS could increase the overall received power and thus the channel propagation gain by up to 3\,dB. This indicates that the implementation of RIS technology can positively influence the propagation environment, and that this can be done at low complexity and low cost.

\bibliographystyle{IEEEtran}
\bibliography{references}

\begin{thebibliography}{10}
\providecommand{\url}[1]{#1}
\csname url@samestyle\endcsname
\providecommand{\newblock}{\relax}
\providecommand{\bibinfo}[2]{#2}
\providecommand{\BIBentrySTDinterwordspacing}{\spaceskip=0pt\relax}
\providecommand{\BIBentryALTinterwordstretchfactor}{4}
\providecommand{\BIBentryALTinterwordspacing}{\spaceskip=\fontdimen2\font plus
\BIBentryALTinterwordstretchfactor\fontdimen3\font minus \fontdimen4\font\relax}
\providecommand{\BIBforeignlanguage}[2]{{%
\expandafter\ifx\csname l@#1\endcsname\relax
\typeout{** WARNING: IEEEtran.bst: No hyphenation pattern has been}%
\typeout{** loaded for the language `#1'. Using the pattern for}%
\typeout{** the default language instead.}%
\else
\language=\csname l@#1\endcsname
\fi
#2}}
\providecommand{\BIBdecl}{\relax}
\BIBdecl

\bibitem{haneda2015statistical}
K.~Haneda, J.~Järveläinen, A.~Karttunen, M.~Kyrö, and J.~Putkonen, ``A statistical spatio-temporal radio channel model for large indoor environments at 60 and 70 {GHz},'' \emph{IEEE Transactions on Antennas and Propagation}, vol.~63, no.~6, pp. 2694--2704, 2015.

\bibitem{fan2016measured}
W.~Fan, I.~Carton, J.~{\O}. Nielsen, K.~Olesen, and G.~F. Pedersen, ``Measured wideband characteristics of indoor channels at centimetric and millimetric bands,'' \emph{EURASIP Journal on Wireless Communications and Networking}, vol. 2016, no.~1, pp. 1--13, 2016.

\bibitem{jiang2023sensing}
S.~Jiang, A.~Hindy, and A.~Alkhateeb, ``Sensing aided reconfigurable intelligent surfaces for {3GPP 5G} transparent operation,'' \emph{IEEE Transactions on Communications}, pp. 1--1, 2023.

\bibitem{ouyang2023computer}
M.~Ouyang, F.~Gao, Y.~Wang, S.~Zhang, P.~Li, and J.~Ren, ``Computer vision-aided reconfigurable intelligent surface-based beam tracking: Prototyping and experimental results,'' \emph{IEEE Transactions on Wireless Communications}, pp. 1--1, 2023.

\bibitem{sanchez2021optimal}
J.~Sanchez, E.~Bengtsson, F.~Rusek, J.~Flordelis, K.~Zhao, and F.~Tufvesson, ``Optimal, low-complexity beamforming for discrete phase reconfigurable intelligent surfaces,'' in \emph{IEEE Global Communications Conference (GLOBECOM)}, 2021, pp. 01--06.

\bibitem{cai2023switched}
X.~Cai, E.~L. Bengtsson, O.~Edfors, and F.~Tufvesson, ``A switched array sounder for dynamic millimeter-wave channel characterization: Design, implementation and measurements,'' \emph{IEEE Transactions on Antennas and Propagation}, 2023.

\bibitem{yezhen2020novel}
L.~Yezhen, R.~Yongli, Y.~Fan, X.~Shenheng, and Z.~Jiannian, ``A novel 28 {GHz} phased array antenna for {5G} mobile communications,'' \emph{ZTE Communications}, vol.~18, no.~3, pp. 20--25, 2020.

\bibitem{HRPEcai}
{X. Cai \textit{et al.}}, ``Enabling complexity-efficient high-resolution parameter estimation for wideband switched array channel sounding,'' \emph{Under preparation}, 2023.

\bibitem{fleury1999channel}
B.~H. Fleury, M.~Tschudin, R.~Heddergott, D.~Dahlhaus, and K.~I. Pedersen, ``Channel parameter estimation in mobile radio environments using the {SAGE} algorithm,'' \emph{IEEE Journal on selected areas in communications}, vol.~17, no.~3, pp. 434--450, 1999.

\bibitem{cai2023enhanced}
X.~Cai, M.~Zhu, A.~Fedorov, and F.~Tufvesson, ``\BIBforeignlanguage{English}{Enhanced effective aperture distribution function for characterizing large-scale antenna arrays},'' \emph{\BIBforeignlanguage{English}{IEEE Transactions on Antennas and Propagation}}, Jun. 2023.

\bibitem{caire1999capacity}
G.~Caire and S.~Shamai, ``On the capacity of some channels with channel state information,'' \emph{IEEE Transactions on Information Theory}, vol.~45, no.~6, pp. 2007--2019, 1999.

\bibitem{vu2007mimo}
M.~Vu and A.~Paulraj, ``{MIMO} wireless linear precoding,'' \emph{IEEE Signal Processing Magazine}, vol.~24, no.~5, pp. 86--105, 2007.

\bibitem{you2020spectral}
L.~You, J.~Xiong, A.~Zappone, W.~Wang, and X.~Gao, ``Spectral efficiency and energy efficiency tradeoff in massive {MIMO} downlink transmission with statistical {CSIT},'' \emph{IEEE Transactions on Signal Processing}, vol.~68, pp. 2645--2659, 2020.

\bibitem{molisch2012wireless}
A.~F. Molisch, \emph{Wireless communications}.\hskip 1em plus 0.5em minus 0.4em\relax John Wiley \& Sons, 2012, vol.~34.

\end{thebibliography}

\end{document}